\documentclass[11pt,a4paper]{article}
\pdfoutput=1

\usepackage[utf8]{inputenc}
\usepackage{amsfonts,amssymb}
\usepackage[colorlinks=0]{hyperref}
\usepackage{sitender}
\usepackage{graphicx}
\usepackage{mathrsfs}
\usepackage{comment}
\usepackage{cite}
\usepackage{cite}
\usepackage{graphicx}
\usepackage{float}
\usepackage[normalem]{ulem}
\usepackage[utf8]{inputenc}
\usepackage[T1]{fontenc}
\usepackage{lmodern, textcomp}
\usepackage[english]{babel}
\usepackage{csquotes,cancel}

\usepackage[left=1in,right=1in,top=.8in,bottom=1in]{geometry}

\usepackage{eurosym, xspace}
\usepackage{graphicx, float}
\usepackage{verbatim}

 \usepackage{upgreek}
\usepackage{amsmath, amsfonts, amssymb,upgreek}
\usepackage{cancel}
\usepackage{mathtools}
\usepackage{slashed}

\usepackage{framed}
\usepackage[amsmath, hyperref, framed]{}
\usepackage{hhline}
\usepackage{bm}

\usepackage{array, caption}
\usepackage{array,multirow}

\newcolumntype{C}[1]{>{\centering\arraybackslash}m{#1}}

\makeatletter
\newsavebox{\@brx}
\newcommand{\llangle}[1][]{\savebox{\@brx}{\(\m@th{#1\langle}\)}%
  \mathopen{\copy\@brx\kern-0.5\wd\@brx\usebox{\@brx}}}
\newcommand{\rrangle}[1][]{\savebox{\@brx}{\(\m@th{#1\rangle}\)}%
  \mathclose{\copy\@brx\kern-0.5\wd\@brx\usebox{\@brx}}}
\makeatother

 \input epsf.sty

\usepackage{bm}

\begin{document}

\baselineskip 24pt

\begin{center}
{\Large \bf Two point amplitude for closed superstrings }

\end{center}

\vskip .6cm
\medskip

\vspace*{4.0ex}

\baselineskip=18pt

\begin{center}

{\large 
\rm  Sitender Pratap Kashyap$^{a,b}$ }

\end{center}

\vspace*{4.0ex}

\centerline{\it \small $^a$Institute of Physics of the Czech Academy of Sciences \& CEICO, }
\centerline{ \it \small Na Slovance 2, 182 21, Prague - Czech Republic}
\centerline{ \it \small $^b$ ICTP - South American Institute for Fundamental Research, IFT-UNESP,}
\centerline{\it \small  Rua Dr. Bento Teobaldo Ferraz 271, 01140-070, S\~ao Paulo, SP, Brazil}

\centerline{\small E-mail:  kashyap@fzu.cz }

\vspace*{5.0ex}

\centerline{\bf Abstract} \bigskip

We present a prescription for computing the tree-level two-point amplitude of closed strings in the
pure spinor superstring formalism, thereby completing the analysis of such superstring amplitudes. The
construction relies on fixing the residual conformal Killing group using a mostly BRST-exact operator
that has been successfully applied in the open-string case. Earlier attempts at a straightforward extension
to closed strings—treating them naïvely as products of open strings—fail. Nevertheless, we show that
a consistent prescription can be obtained by replacing the open-string BRST charge with the closed-
string BRST charge. The key idea is to employ closed-string vertex operators with nonstandard ghost-
number assignments, rather than the conventional ghost-number (1, 1) vertices. Furthermore, since the
pure spinor BRST cohomology for closed strings vanishes at total (left plus right) ghost number four or
higher, we find that the resulting prescription is essentially unique.

\vfill

\vfill \eject

\baselineskip18pt

\tableofcontents

\section{Introduction}

Sometimes, a highly convincing but ultimately hand-waving argument can obscure subtle details, leading to faulty conclusions and incorrect physical interpretations. A notable example of this phenomenon in the string theory literature concerns the tree-level two-point amplitude. Traditionally, an argument based on the infinite volume of the conformal Killing group (CKG) was used to claim that amplitudes with fewer than three vertex operators on genus-zero surfaces vanish. As a consequence, two-point amplitudes were long assumed to be zero, leading to the inference that the Polyakov prescription computes only the connected part of the scattering amplitude rather than the full S-matrix. In other words, when one writes $S=1+iT$, string amplitudes were taken—by analogy with conventional quantum field theory—to compute only the $iT$ part rather than the full $S$.

This viewpoint was rigorously revisited in \cite{Erbin:2019uiz} for the case of bosonic strings, where a flaw in the earlier arguments was identified and the finiteness of these two-point amplitudes was demonstrated. This revised understanding is fully consistent with the requirements of cluster decomposition, unlike the earlier interpretation. Since the fixing of the bosonic CKG carries over in a straightforward manner to the Ramond–Neveu–Schwarz (RNS) formalism of superstrings, the tree-level two-point amplitudes must also be finite in superstring theory.

There exist two additional, equivalent formalisms for quantizing superstrings, namely the Green–Schwarz (GS) and the Pure Spinor (PS) formalism due to Berkovits. The GS formalism suffers from significant quantization difficulties arising from the complicated nature of its constraints and is therefore largely avoided in practical computations. In contrast, the pure spinor formalism—the only manifestly super-Poincaré covariant method for quantizing superstrings—has established itself as a powerful and consistent framework for both amplitude computations and the quantization of strings in curved backgrounds. It has not only reproduced all non-trivial amplitudes obtained in the RNS formalism, but in many respects has surpassed it. However, the pure spinor formalism is formulated in a language entirely different from that of RNS, and consequently the results of \cite{Erbin:2019uiz} cannot be directly lifted to this setting. It therefore remains an important open issue to demonstrate explicitly that even the simplest amplitudes are finite in the pure spinor formalism, in order to preserve its equivalence with the RNS formalism. Motivated by this gap, we investigated these amplitudes in \cite{Kashyap:2020tgx} and established finiteness in the case of open strings.
At first sight, closed strings—being naïvely viewed as a left–right product of open strings—might be expected to lead to the same conclusions. However, a closer examination reveals that this expectation is incorrect, and that the case of closed strings requires special attention.

It is useful to recall at this stage that there exist several equivalent quantization schemes for bosonic strings. The analysis of \cite{Erbin:2019uiz} employed the Faddeev–Popov covariant gauge-fixing method, but the same physical conclusions must also emerge in other quantization schemes, such as the operator formalism. This issue was addressed for open bosonic strings in the operator formalism in \cite{Seki:2019ycz}, and for closed bosonic strings in the BRST formalism in \cite{Kishimoto:2024aig}. Both studies further demonstrate that it is naïve to regard closed strings as a simple product of two open strings. An alternative proof based on Liouville regularization was provided in \cite{Giribet:2023gub}. These works make it clear that the details underlying the finiteness of two-point amplitudes depend sensitively on the formalism under consideration.

Let us now return to the pure spinor formalism. Unlike the bosonic string, there is no known worldsheet reparametrization-invariant action whose gauge fixing can be used to provide a first-principles derivation of the finiteness of these amplitudes\footnote{see however, \cite{Jusinskas:2019vmd} for attempts in this direction}. The amplitude prescription in the pure spinor formalism is instead inspired by that of the bosonic string, as the pure spinor formalism is closely related to the $N=2$ topological string \cite{Berkovits_topological}, which inherits the amplitude prescription from the bosonic strings. Thus, even in the absence of a first-principles derivation, a consistent amplitude prescription exists in the pure spinor formalism \cite{Berkovits}\footnote{see \cite{Hoogeveen:2007tu} for an alternative derivation}. However, this prescription implicitly assumes that the conformal Killing group on genus-zero surfaces has been completely fixed and is therefore applicable only to amplitudes involving three or more external states. A naïve extension to the two-point function leads to a vanishing result (see section \ref{review}), which cannot be correct—both for consistency with the RNS formalism and for conceptual reasons tied to cluster decomposition. This necessitates the search for an alternative prescription for two-point amplitudes in the pure spinor formalism.

Our approach to establishing the finiteness of these amplitudes in the pure spinor formalism is inspired by the proofs given in \cite{Seki:2019ycz,Kishimoto:2024aig}. A crucial ingredient in these works is the use of a mostly BRST (mBRST)-exact operator to saturate the $c$-ghost zero modes. Since pure spinor scattering amplitudes inherit much of their structure from bosonic string theory, we similarly employ an mBRST-exact operator to demonstrate the finiteness of two-point amplitudes in the pure spinor formalism. In \cite{Kashyap:2022qtd}, such an operator was derived from a Faddeev–Popov gauge fixing of the bosonic string path integral, thereby showing as expected, equivalence of the path integral and the operator methods. Further, it was shown to provide a valid alternative gauge-fixing prescription even in pure spinor amplitudes, for other tree-level amplitudes as well.

In \cite{Kashyap:2020tgx}, this idea was implemented for open strings using the following m-BRST-exact operator
\be
V_0(z,\bar{z})\equiv\int dq; (\l\g^0\t) ;e^{iqX^0(z,\bar{z})} \label{mBRST_exact_open_string}
\ee
where $q\in\mathbb{R}$ is a parameter. For $q\neq 0$, the integrand can be re-expressed as a BRST-exact operator $[Q,e^{iqX^0}]$. Here $Q=\oint dz, \l^\a d_\a$ is the BRST charge of the pure spinor formalism, with $\l^\a$ a commuting spinor satisfying $\l^\a\rb{\g^m}_{\a\b}\l^\b=0$ for all gamma matrices $\g^m$. The fields $(X^m,\t^\a)$ form the standard $\mc{N}=1$ superspace, with $d_\a$ denoting the supersymmetrized conjugate momentum of $\t^\a$. A brief summary of the pure spinor formalism is provided in section \ref{review}.

With this ingredient, the open-string two-point amplitude prescription reads
\be
\mc{A}_2=\la V_0(z_0) V_1(z_1) V(z_2)\ra \label{norm_open}
\ee
where $V_1$ and $V_2$ are vertex operators of ghost number one. This correlator is normalized using\footnote{An alternative normalization based on zero $\l^\a$ was proposed in \cite{Berkovits:2016xnb}. In principle, this could have been used to define an alternative two-point amplitude prescription. However, it was recently shown that this prescription is incorrect \cite{Azevedo:2025con}. A naïve generalization works only for bosons and fails for fermions, in agreement with the findings of \cite{Azevedo:2025con}.}
\be
\Big\la \rb{\l \g^m \t} \rb{\l \g^n \t} \rb{\l \g^p \t} \rb{\t \g_{mnp} \t} \Big\ra=1
\ee
where the additional four $\t^\a$ arise from the $\theta$-expansion of $V_1$ and $V_2$.
A naïve generalization of this prescription to closed strings fails. In this case, $Q_{\text{open}}\rightarrow Q_{\text{closed}}$, and $V_1$ and $V_2$ are closed-string vertex operators of ghost number $(1,1)$. This failure is evident from the fact that \eqref{norm_open} generalizes to
\be
\Big\la \rb{\l \g^m \t} \rb{\l \g^n \t} \rb{\l \g^p \t} \rb{\t \g_{mnp} \t} \rb{\hat{\l} \g^s \hat{\t}} \rb{\hat{\l} \g^t \hat{\t}} \rb{\hat{\l} \g^u \hat{\t}} \rb{\hat{\l} \g_{stu} \hat{\t}} \Big\ra =1 \label{norm_closed}
\ee
where hatted quantities denote the right-moving sector. For standard unintegrated vertex operators of ghost number $(1,1)$, it is impossible to satisfy this zero-mode normalization: if the left-moving sector supplies the required three $\l^\a$, the right-moving sector supplies only two $\hat{\l}^\a$, and vice-versa.

In this work, we show that one can, nonetheless, continue to employ the mBRST-exact operator for closed strings,
\be
V_0(z,\bar{z})\equiv\int dq \sqb{(\l\g^0\t)+ (\hat{\l}\g^0\hat{\t})}e^{iqX^0(z,\bar{z})} \label{mBRST_exact}
\ee
while making use of unintegrated vertex operators with different ghost numbers in order to obtain a finite result. As before, this operator is called mBRST-exact since, for $q\neq 0$, the integrand can be written as
\be
\sqb{(\l\g^0\t)+ (\hat{\l}\g^0\hat{\t})}e^{iqX^0(z,\bar{z})}\propto \{Q_{closed},e^{iqX^0}\}
\ee
where $Q_{closed}=Q_L+Q_R=Q+\bar{Q}$ is the sum of the BRST charges associated with the holomorphic and anti-holomorphic sectors.

In this work we present a consistent and essentially unique prescription for computing the tree-level two-point amplitude of closed superstrings in the pure spinor formalism. The paper is organized as follows. In section \ref{review} we briefly review the aspects of the pure spinor formalism relevant for this work. In section \ref{vertex} we discuss vertex operators with varying ghost numbers. The amplitude prescription is presented in section \ref{prescription}, and we conclude with a discussion in section \ref{discuss}.

\section{Pure Spinor review} \label{review}
The pure spinor formalism can be formulated as a worldsheet conformal field theory in the conformal gauge, with action (displaying only the holomorphic sector) \cite{Berkovits}
\be
S=\int d^2z \rb{\p X^m \bar\p X_m + p_\a \bar\p\t^\a +\l^a\bar \p w_\a} , 
\ee
Here $(X^m,\t^\a)$ parametrize an $\mc{N}=1$ superspace in flat ten-dimensional spacetime, where $X^m$ denotes the spacetime coordinate and $\t^\a$ is an anticommuting Weyl fermionic coordinate. The field $\l^\a$ is a commuting Weyl spinor subject to the pure spinor constraint $\l\g^m\l=0$ for all gamma matrices $\g^m$. Its conjugate momentum $w_\a$ possesses a gauge invariance inherited from this constraint.
The pure spinor $\l^\a$ and the supersymmetric constraint $d_\a$ are used to construct the BRST charge of the formalism,
\be
Q=\int dz\, \l^\a d_\a ,
\ee
where $d_\a$ is the supersymmetric conjugate momentum of $\theta^\alpha$. Correlation functions are normalized by fixing the zero-mode integral of the fields according to
\be
\la \rb{\l\g^m\t} \rb{\l\g^n\t} \rb{\l\g^p\t} \rb{\t\g_{mnp}\t}\ra =1 \label{norm_open_corr}
\ee
which plays a central role in determining which correlators are non-vanishing. Physical vertex operators are defined as elements of the BRST cohomology at ghost number one, although—as we shall review—equivalent descriptions exist at other ghost numbers.
\subsection{The Amplitude prescription}
In the pure spinor formalism, scattering amplitudes are computed using a prescription that involves both unintegrated and integrated vertex operators \cite{Berkovits},
\be
\mc{A}_n=\int \prod_{i=4}^n dz_i\la V(z_1,\bar{z}_1) V(z_2,\bar{z}_2) V(z_3,\bar{z}_3) U(z_4,\bar{z}_4) \cdots U(z_n,\bar{z}_n)\ra \label{amplitude_pes}
\ee
where $V$ denotes an unintegrated vertex operator and $U$ its integrated counterpart.
An alternative prescription, involving only integrated vertex operators, was proposed in \cite{Berkovits:2016xnb},
\be
\mc{A}_n= \la c(z_1,\bar{z}_1) c(z_2,\bar{z}_2) c(z_3,\bar{z}_3) U(z_1,\bar{z}_1) U(z_2,\bar{z}_2) U(z_3,\bar{z}_3) \int dz_4 d\bar{z}_4 U(z_4,\bar{z}_4) \cdots \int dz_n d\bar{z}_n U(z_n,\bar{z}_n)\ra.
\ee
However, it was recently shown in \cite{Azevedo:2025con} that this second prescription is incorrect, as it fails to reproduce even the simplest three-point super Yang–Mills amplitude. We will therefore not consider it further.
It is important to emphasize that the prescription \eqref{amplitude_pes} already assumes the presence of three unintegrated vertex operators. As a result, it cannot be directly applied to the computation of two-point amplitudes. A naïve generalization would involve only two unintegrated vertex operators and consequently yields a vanishing result due to the zero-mode normalization condition \eqref{norm_open_corr}. Since it is known that tree-level two-point amplitudes in bosonic string theory are finite, consistency demands the existence of an alternative prescription that yields a finite two-point amplitude in the pure spinor formalism as well. Establishing such a prescription has been the objective of \cite{Kashyap:2020tgx,Kashyap:2022qtd} and is also the focus of the present work.

\section{Vertex operators in various ghost numbers} \label{vertex}
A systematic study of vertex operators for closed strings was initiated in \cite{Grassi_closed_vertex}. The authors introduced a descent procedure that relates integrated and unintegrated vertex operators corresponding to physical states in the cohomology of the pure spinor BRST charge for closed strings. Specifically, they define
\be
O^{(1,1)}_{0,0}=\mc{V}^{(1,1)},\quad O^{(0,1)}_{0,1}=\mc{V}^{(0,1)}d\bar{z}, \quad O^{(1,0)}_{1,0}=\mc{V}^{(1,0)} dz, \quad O^{(0,0)}_{1,1}=\mc{V}^{(0,0)} dz \wedge d\bar{z},
\ee
where the operators satisfy the following descent relations:
\be
[Q_L, O^{(a,b)}_{c,d}\} = \p O^{(a+1,b)}{c-1,d}, \quad [Q_R, O^{(a,b)}_{c,d}\} = \bar{\p} O^{(a,b+1)}{c,d-1}.
\ee
Here, $[\cdots, *\}$ denotes either a commutator or an anticommutator depending on the Grassmann parity of the operators involved. The total ghost number of an operator $O^{(a,b)}$ is defined as $a+b$.
The analysis in \cite{Grassi_closed_vertex} focused primarily on unintegrated vertex operators of ghost number two and their corresponding integrated operators. In this notation, a general ghost number two vertex operator can be written as
\be
V^2 = V^{(2,0)} + V^{(1,1)} + V^{(0,2)},
\ee
while the operator defined in \eqref{mBRST_exact} can be expressed as
\be
V_0 = V_0^{(1,0)} + V_0^{(0,1)}.
\ee
The pure spinor cohomology at other ghost numbers was explored in \cite{Mikhailov:2012uh}, where it was shown that it vanishes for ghost number four or higher. Furthermore, \cite{Mikhailov:2014qka} demonstrated that the cohomology at ghost number three is non-trivial and, in fact, equivalent to that at ghost number two. Using the notation introduced above, the ghost number three vertex operator in type IIB string theory can be written as \cite{Mikhailov:2014qka}
\be
V_3\equiv \rb{a_m(\l\g^m\t)-a_m(\hat{\l}\g^m\hat{\t})}V_2,
\ee
where $a_m$ is a constant vector satisfying $a_m k^m\ne 0$, and $V_2$ is given by
\be
\Big[(\l\g^m\t)(\g_m\t)\a+[\l\t^{\ge 4}]\a\Big] P^{\a\hat{\b}}\Big[(\hat{\l}\g^m\hat{\t})(\g_m\hat{\t}){\hat{\b}}+[\hat{\l}\hat{\t}^{\ge 4}]{\hat{\b}}\Big] e^{ik.X}.
\ee
Here, $P^{\a\hat{\b}}$ is a constant polarization tensor satisfying $k^m(\g_m){\a\b}P^{\b\hat{\b}}=0=P^{\b\hat{\b}}(\g^m){\hat{\b}\hat{\a}}k_m$, and the terms in $\sqb{\cdot}$ denote contributions of order four or higher in $\t$ and $\hat{\t}$.
The fact that the only non-trivial pure spinor cohomology beyond ghost number two occurs at ghost number three is crucial for our analysis and underlies the uniqueness of the amplitude prescription, which constitutes the main result of this work.

\section{The new amplitude prescription} \label{prescription}

We now show that by employing ghost number three vertex operators, it is possible to obtain a non-vanishing tree-level two-point function for closed strings using the mBRST-exact operator introduced in \eqref{mBRST_exact}. Before doing so, however, it is instructive to recall why the two-point function vanishes when one uses ghost number two vertex operators together with the same mBRST-exact operator.
To this end, we recall that a closed-string correlator in the pure spinor formalism is normalized as
\be
\Big\la \rb{\l \g^m \t} \rb{\l \g^n \t} \rb{\l \g^p \t} \rb{\t \g_{mnp} \t} \rb{\hat{\l} \g^s \hat{\t}} \rb{\hat{\l} \g^t \hat{\t}} \rb{\hat{\l} \g^u \hat{\t}} \rb{\hat{\l} \g_{stu} \hat{\t}} \Big\ra =1 \label{norm_corr}
\ee
which enforces the saturation of three pure spinor zero modes in both the left- and right-moving sectors.
Let us consider the correlator
\be
\mc{M}_2&=&\Big\la V_0(z_0,\bar{z}_0) V^{2}(z_1,\bar{z}_1) V^{2}(z_2,\bar{z}2) \Big\ra\non\\
&=&\int dq \Big\la \rb{V_0^{(1,0)}+ V{0}^{(0,1)}}(z_0,\bar{z}_0) \rb{V^{(2,0)} + V^{(1,1)} + V^{(0,2)}}(z_1,\bar{z}_1) \rb{V^{(2,0)} + V^{(1,1)} + V^{(0,2)}}(z_2,\bar{z}_2) \Big\ra\non\\
&\propto&\la \mc{V}^{(1+2+2,0+0+0)}\ra + \cdots+ \la \mc{V}^{(0+0+0,1+2+2)}\ra\non\\
&=&\la \mc{V}^{(5,0)}\ra + \cdots+ \la \mc{V}^{(0,5)}\ra \label{trial_corr}
\ee
where $\mc{V}^{(i,j)}$ denotes an operator carrying ghost number $i$ in the left-moving sector and $j$ in the right-moving sector. Crucially, we observe that no term with ghost numbers $(i,j)=(3,3)$ appears in \eqref{trial_corr}. Consequently, all contributions vanish as a direct result of the normalization condition \eqref{norm_corr}.
The above argument also allows us to determine the conditions under which a non-vanishing correlator can arise. In order to saturate the zero modes and produce a non-trivial correlator, the total ghost number must be six, with three units in the left-moving sector and three in the right-moving sector. For example, if both vertex operators were taken to have ghost number three, the total ghost number in \eqref{trial_corr} would be seven, which again fails to produce a non-vanishing result. It follows that the only viable possibility is to take one vertex operator of ghost number three and the other of ghost number two. The remaining ambiguity lies in which operator is assigned which ghost number.
For definiteness, we choose an ordering in which the ghost numbers increase and define
\be
\mc{A}_2&\equiv&\Big\la V_0(z_0,\bar{z}_0) V^{2}(z_1,\bar{z}_1) V^{3}(z_2,\bar{z}_2) \Big\ra\non\\
&=&\int dq \Big\la \rb{V_0^{(1,0)}+ V{0}^{(0,1)}} \rb{V^{(2,0)} + V^{(1,1)} + V^{(0,2)}} \rb{V^{(3,0)}+ V^{(2,1)} + V^{(1,2)} + V^{(0,3)}} \Big\ra \non\\
\ee
It is then straightforward to verify that this prescription yields the following non-vanishing correlators,
\be
&&\la V_0^{(1,0)} V^{(2,0)} V^{(0,3)}\ra \;,\;
\la V_0^{(1,0)} V^{(1,1)} V^{(1,2)}\ra \;,\;
\la V_0^{(1,0)} V^{(0,2)} V^{(2,1)}\ra\non\\
&& \la V_0^{(0,1)} V^{(2,0)} V^{(1,2)}\ra \;,\;
\la V_0^{(0,1)} V^{(3,0)} V^{(2,1)}\ra \;,\;
\la V_0^{(0,1)} V^{(0,2)} V^{(2,1)}\ra 
\ee
while the remaining eighteen combinations vanish identically. Together with the fact that the BRST cohomology at ghost number four or higher is trivial, this demonstrates that the above choice is unique in yielding a non-vanishing tree-level two-point correlator. Following the arguments presented in \cite{Kashyap:2020tgx}, we find that $\mc{A}_2$ reproduces the expected kinematic dependence of the two-point function.

\section{Discussion} \label{discuss}
We have shown that the mBRST-exact operator can be used to compute the tree-level two-point amplitude in the closed pure spinor superstring, provided one allows vertex operators of different ghost numbers. This fills a long-standing gap in the literature by demonstrating that tree-level two-point superstring amplitudes are non-vanishing and reproduce the standard field-theory expectation. In particular, this result supports the interpretation that the Polyakov path integral computes the full string S-matrix, in a manner consistent with cluster decomposition.

At first sight, our construction appears to rely on a special choice of spacetime coordinate $X^0$, potentially raising concerns about Lorentz invariance. However, this apparent asymmetry is only superficial. As shown in \cite{Kashyap:2022qtd}, the vertex operator can be made manifestly covariant by introducing a time-like vector satisfying $t^2=-1$. The BRST invariance and super-Poincaré invariance of the resulting prescription then follow along the same lines as in the open-string analysis of \cite{Kashyap:2020tgx}.

Beyond their conceptual significance, these amplitudes provide a practical method for fixing the overall normalization of vertex operators through factorization of two-point amplitudes, rather than relying on four-point amplitudes. It would be interesting to explore whether mBRST-exact operators can also be employed to gauge-fix the conformal Killing vectors in one-loop amplitudes. Taken together, our results further solidify the equivalence between the pure spinor and RNS formalisms and clarify the role of two-point amplitudes in the consistent definition of the superstring S-matrix.

\section*{Acknowledgments}
I am grateful to Mritunjay for useful comments on an earlier draft of this manuscript. I am supported by the Marie Sklodowska-Curie Actions -- COFUND project, which is co-funded by the European Union (Physics for Future -- Grant Agreement No. 101081515). I also thank ICTP-SAIFR (FAPESP grant 2021/14335-0), where part of this work was carried out.

\bibliographystyle{JHEP}
\bibliography{reference}

@article{Berkovits:2016xnb,
    author = "Berkovits, Nathan",
    title = "{Untwisting the pure spinor formalism to the RNS and twistor string in a flat and AdS$_{5} \times$ S$^{5}$ background}",
    eprint = "1604.04617",
    archivePrefix = "arXiv",
    primaryClass = "hep-th",
    doi = "10.1007/JHEP06(2016)127",
    journal = "JHEP",
    volume = "06",
    pages = "127",
    year = "2016"
}

@article{Azevedo:2025con,
    author = "Azevedo, Thales and Georgoudis, Alessandro and Lipinski Jusinskas, Renann and Kashyap, Sitender Pratap",
    title = "{Tension between string amplitude prescriptions with presumed spacetime supersymmetry}",
    eprint = "2509.15111",
    archivePrefix = "arXiv",
    primaryClass = "hep-th",
    reportNumber = "QMUL-PH-25-26",
    doi = "10.1007/JHEP11(2025)118",
    journal = "JHEP",
    volume = "11",
    pages = "118",
    year = "2025"
}

@article{Jusinskas:2019vmd,
    author = "Jusinskas, Renann Lipinski",
    title = "{Towards the underlying gauge theory of the pure spinor superstring}",
    eprint = "1903.10753",
    archivePrefix = "arXiv",
    primaryClass = "hep-th",
    doi = "10.1007/JHEP10(2019)063",
    journal = "JHEP",
    volume = "10",
    pages = "063",
    year = "2019"
}

@article{Kishimoto:2024aig,
    author = "Kishimoto, Isao and Seki, Shigenori and Takahashi, Tomohiko",
    title = "{Two-point closed string amplitudes in the BRST formalism}",
    eprint = "2402.07464",
    archivePrefix = "arXiv",
    primaryClass = "hep-th",
    doi = "10.1016/j.physletb.2024.138657",
    journal = "Phys. Lett. B",
    volume = "853",
    pages = "138657",
    year = "2024"
}

@article{Giribet:2023gub,
    author = "Giribet, Gaston and Labranche, Nicholas and La Madrid, Joan",
    title = "{Remarks on the two-point string amplitudes}",
    eprint = "2303.15658",
    archivePrefix = "arXiv",
    primaryClass = "hep-th",
    doi = "10.1103/PhysRevD.107.106021",
    journal = "Phys. Rev. D",
    volume = "107",
    number = "10",
    pages = "106021",
    year = "2023"
}

@article{Kashyap:2020tgx,
    author = "Kashyap, Sitender Pratap",
    title = "{Two-point superstring tree amplitudes using the pure spinor formalism}",
    eprint = "2012.03802",
    archivePrefix = "arXiv",
    primaryClass = "hep-th",
    doi = "10.21468/SciPostPhysCore.8.1.005",
    journal = "SciPost Phys. Core",
    volume = "8",
    pages = "005",
    year = "2025"
}

@article{Kashyap:2022qtd,
    author = "Kashyap, Sitender Pratap",
    title = "{The mostly BRST exact operator in superstrings}",
    eprint = "2212.13838",
    archivePrefix = "arXiv",
    primaryClass = "hep-th",
    doi = "10.1007/JHEP02(2025)129",
    journal = "JHEP",
    volume = "02",
    pages = "129",
    year = "2025"
}

@article{Mikhailov:2012uh,
    author = "Mikhailov, Andrei",
    title = "{Pure spinors in AdS and Lie algebra cohomology}",
    eprint = "1207.2441",
    archivePrefix = "arXiv",
    primaryClass = "hep-th",
    doi = "10.1007/s11005-014-0705-2",
    journal = "Lett. Math. Phys.",
    volume = "104",
    pages = "1201--1233",
    year = "2014"
}

@article{Mikhailov:2014qka,
    author = "Mikhailov, Andrei",
    title = "{Vertex operators of ghost number three in Type IIB supergravity}",
    eprint = "1401.3783",
    archivePrefix = "arXiv",
    primaryClass = "hep-th",
    doi = "10.1016/j.nuclphysb.2016.04.007",
    journal = "Nucl. Phys. B",
    volume = "907",
    pages = "509--541",
    year = "2016"
}

@article{Berkovits,
      author         = "Berkovits, Nathan",
      title          = "{Super Poincare covariant quantization of the
                        superstring}",
      journal        = "JHEP",
      volume         = "04",
      year           = "2000",
      pages          = "018",
      doi            = "10.1088/1126-6708/2000/04/018",
      eprint         = "hep-th/0001035",
      archivePrefix  = "arXiv",
      primaryClass   = "hep-th",
      reportNumber   = "IFT-P-005-2000",
      SLACcitation   = "%%CITATION = HEP-TH/0001035;%%"
}

@article{Hoogeveen:2007tu,
    author = "Hoogeveen, Joost and Skenderis, Kostas",
    title = "{BRST quantization of the pure spinor superstring}",
    eprint = "0710.2598",
    archivePrefix = "arXiv",
    primaryClass = "hep-th",
    reportNumber = "ITFA-2007-49",
    doi = "10.1088/1126-6708/2007/11/081",
    journal = "JHEP",
    volume = "11",
    pages = "081",
    year = "2007"
}

@article{Berkovits_topological,
      author         = "Berkovits, Nathan",
      title          = "{Pure spinor formalism as an N=2 topological string}",
      journal        = "JHEP",
      volume         = "10",
      year           = "2005",
      pages          = "089",
      doi            = "10.1088/1126-6708/2005/10/089",
      eprint         = "hep-th/0509120",
      archivePrefix  = "arXiv",
      primaryClass   = "hep-th",
      reportNumber   = "IFT-P-031-2005",
      SLACcitation   = "%%CITATION = HEP-TH/0509120;%%"
}

@article{Grassi_closed_vertex,
      author         = "Grassi, P. A. and Tamassia, L.",
      title          = "{Vertex operators for closed superstrings}",
      journal        = "JHEP",
      volume         = "07",
      year           = "2004",
      pages          = "071",
      doi            = "10.1088/1126-6708/2004/07/071",
      eprint         = "hep-th/0405072",
      archivePrefix  = "arXiv",
      primaryClass   = "hep-th",
      reportNumber   = "YITP-SB-04-12, IHES-P-04-20, FNT-T-2004-04",
      SLACcitation   = "%%CITATION = HEP-TH/0405072;%%"
}

@article{Erbin:2019uiz,
      author         = "Erbin, Harold and Maldacena, Juan and Skliros, Dimitri",
      title          = "{Two-Point String Amplitudes}",
      journal        = "JHEP",
      volume         = "07",
      year           = "2019",
      pages          = "139",
      doi            = "10.1007/JHEP07(2019)139",
      eprint         = "1906.06051",
      archivePrefix  = "arXiv",
      primaryClass   = "hep-th",
      SLACcitation   = "%%CITATION = ARXIV:1906.06051;%%"
}

@article{Seki:2019ycz,
      author         = "Seki, Shigenori and Takahashi, Tomohiko",
      title          = "{Two-point String Amplitudes Revisited by Operator
                        Formalism}",
      year           = "2019",
      eprint         = "1909.03672",
      archivePrefix  = "arXiv",
      primaryClass   = "hep-th",
      SLACcitation   = "%%CITATION = ARXIV:1909.03672;%%"
}

\end{document}